\documentclass[aps,prc,superscriptaddress,showpacs,amsfonts,twocolumn,floatfix]{revtex4}
\usepackage{graphics}
\usepackage{latexsym}
\usepackage{subfigure,wrapfig}
\usepackage{epsfig,color,rotating,amsmath,delarray,array}
\usepackage{makeidx,pifont,float,amssymb}
\definecolor{gray01}{gray}{0.9}
\definecolor{gray02}{gray}{0.8}
\definecolor{gray03}{gray}{0.7}
\definecolor{gray04}{gray}{0.6}
\definecolor{gray05}{gray}{0.5}
\definecolor{gray06}{gray}{0.4}
\definecolor{gray07}{gray}{0.3}
\definecolor{gray08}{gray}{0.2}
\definecolor{gray09}{gray}{0.1}

\begin{document}

\title{Measurement of the Beam Asymmetry~$\Sigma$ in the Forward Direction for $\vec{\gamma} p\to p\pi^0$}

\newcommand*{\HISKP}{Helmholtz-Institut f\"ur Strahlen- und Kernphysik, Universit\"at Bonn, D-53115 Bonn, Germany}
\newcommand*{\GATCHINA}{Petersburg Nuclear Physics Institute, RU-188350 Gatchina, Russia}
\newcommand*{\PI}{Physikalisches Institut, Universit\"at Bonn, D-53115 Bonn, Germany}
\newcommand*{\KVI}{KVI, 9747 AA Groningen, The Netherlands}
\newcommand*{\FSU}{Department of Physics, Florida State University, Tallahassee, Florida 32306, USA}
\newcommand*{\GIESSEN}{II. Physikalisches Institut, Universit\"at Gie\ss en, D-35392 Gie\ss en, Germany}
\newcommand*{\BASEL}{Physikalisches Institut, Universit\"at Basel, CH-4056 Basel, Switzerland}
\newcommand*{\JUELICH}{Institut f\"ur Kernphysik, Forschungszentrum J\"ulich, D-52428 J\"ulich, Germany}

\author{N.~Sparks} \affiliation{\FSU}
\author{V.~Crede} \affiliation{\FSU}
\author{A.V.~Anisovich} \affiliation{\HISKP}\affiliation{\GATCHINA}
\author{J.C.S.~Bacelar} \affiliation{\KVI}
\author{R.~Bantes} \affiliation{\HISKP}
\author{O.~Bartholomy} \affiliation{\HISKP}
\author{D.~Bayadilov} \affiliation{\HISKP}\affiliation{\GATCHINA}
\author{R.~Beck} \affiliation{\HISKP}
\author{Y.A.~Beloglazov} \affiliation{\GATCHINA}
\author{R.~Castelijns} \affiliation{\KVI}
\author{D.~Elsner} \affiliation{\PI}
\author{R.~Ewald} \affiliation{\PI}
\author{F.~Frommberger} \affiliation{\PI}
\author{Chr.~Funke} \affiliation{\HISKP}
\author{R.~Gregor} \affiliation{\GIESSEN}
\author{A.~Gridnev} \affiliation{\GATCHINA}
\author{E.~Gutz} \affiliation{\HISKP}
\author{W.~Hillert} \affiliation{\PI}
\author{P.~Hoffmeister} \affiliation{\HISKP}
\author{I.~Jaegle} \affiliation{\BASEL}
\author{J.~Junkersfeld} \affiliation{\HISKP}
\author{H.~Kalinowsky} \affiliation{\HISKP}
\author{S.~Kammer} \affiliation{\PI}
\author{Frank~Klein} \affiliation{\PI}
\author{Friedrich~Klein} \affiliation{\PI}
\author{E.~Klempt} \affiliation{\HISKP}
\author{M.~Kotulla} \affiliation{\BASEL}\affiliation{\GIESSEN}
\author{B.~Krusche} \affiliation{\BASEL}
\author{H.~L\"ohner} \affiliation{\KVI}
\author{I.V.~Lopatin} \affiliation{\GATCHINA}
\author{S.~Lugert} \affiliation{\GIESSEN}
\author{D.~Menze} \affiliation{\PI}
\author{T.~Mertens} \affiliation{\BASEL}
\author{J.G.~Messchendorp} \affiliation{\KVI}\affiliation{\GIESSEN}
\author{V.~Metag} \affiliation{\GIESSEN}
\author{M.~Nanova} \affiliation{\GIESSEN}
\author{V.A.~Nikonov} \affiliation{\HISKP}\affiliation{\GATCHINA}
\author{D.~Novinski} \affiliation{\GATCHINA}
\author{R. Novotny} \affiliation{\GIESSEN}
\author{M.~Ostrick} \affiliation{\PI}
\author{L.M. Pant} \affiliation{\GIESSEN}
\author{H.~van~Pee} \affiliation{\HISKP}
\author{M.~Pfeiffer} \affiliation{\GIESSEN}
\author{A.~Roy} \affiliation{\GIESSEN}
\author{A.V.~Sarantsev} \affiliation{\HISKP}\affiliation{\GATCHINA}
\author{S.~Schadmand} \affiliation{\JUELICH}
\author{C.~Schmidt} \affiliation{\HISKP}
\author{H.~Schmieden} \affiliation{\PI}
\author{B.~Schoch} \affiliation{\PI}
\author{S.~Shende} \affiliation{\KVI}
\author{V.~Sokhoyan} \affiliation{\HISKP}
\author{A.~S{\"u}le} \affiliation{\PI}
\author{V.V.~Sumachev} \affiliation{\GATCHINA}
\author{T.~Szczepanek} \affiliation{\HISKP}
\author{U.~Thoma} \affiliation{\HISKP}
\author{D.~Trnka} \affiliation{\GIESSEN}
\author{R.~Varma} \affiliation{\GIESSEN}
\author{D.~Walther} \affiliation{\HISKP}\affiliation{\PI}
\author{Ch.~Wendel} \affiliation{\HISKP}
\author{A.~Wilson} \affiliation{\FSU}
\collaboration{The CBELSA/TAPS Collaboration} \noaffiliation

\begin{abstract}
Photoproduction of neutral pions has been studied with the CBELSA/TAPS detector 
for photon energies between 0.92 and 1.68~GeV at the electron accelerator ELSA. 
The beam asymmetry~$\Sigma$ has been extracted for $115^\circ < \theta_{\rm c.m.} 
< 155^\circ$ of the $\pi^0$~meson and for $\theta_{\rm c.m.} < 60^\circ$. The new beam 
asymmetry data improve the world database for photon energies above 1.5~GeV and, 
by covering the very forward region, extend previously published data for the same 
reaction by our collaboration. The angular dependence of $\Sigma$ shows 
overall good agreement with the SAID parameterization.
\end{abstract}

\date{Received: \today / Revised version:}


\maketitle

\section{\label{Section:Introduction}Introduction}
Baryon resonances exhibit a rich excitation spectrum due to their complicated
substructure. The understanding of this structure of the nucleon and its excited states 
is one of the key questions in hadronic physics. Although most quark models based
on three constituent quark degrees of freedom can describe ground-state baryons 
well, they fail in some important details. Known as the missing-resonance 
problem, these quark models have predicted many more excited states at and above 
2~GeV/$c^2$ than have been found experimentally. Of particular importance are the 
measurements of polarization observables in addition to the extraction of total
and differential cross sections. These polarization observables can be sensitive 
to interference terms in the theoretical interpretation of the data and, thus, can
provide access to otherwise small resonance contributions. The beam asymmetry
$\Sigma$, for example, which arises from a linearly polarized photon beam, addresses 
the non-spin-flip terms in the transition current (e.g., convection currents and 
double spin-flip contributions), whereas spin-flip contributions are projected out 
by a circularly polarized photon beam.

Since the $\pi$~meson has isospin $I=1$, both nucleon resonances $(I=1/2)$ 
and $\Delta$~resonances $(I=3/2)$ can contribute to $\pi^0$~photoproduction off the 
proton. The total $\pi^0$~cross section exhibits three clear peaks and a broad
enhancement around $W\approx 1900$~MeV/$c^2$, which represent the four known resonance 
regions below 2~GeV/$c^2$. The first resonance region below 1500~MeV/$c^2$ is dominated 
by the well-known $\Delta(1232)P_{33}$ resonance with very small contributions of the 
$N(1440)P_{11}$ Roper resonance. The $N(1520)D_{13}$ and the two $S_{11}$ resonances 
combined, $N(1535)S_{11}$ and $N(1650)S_{11}$, contribute with about equal strength 
to the second resonance region around 1550~MeV/$c^2$. The third bump in the $p\pi^0$
total cross section is mainly due to three major resonance contributions: $\Delta(1700)
D_{33}$, $N(1680)F_{15}$, and $N(1650)S_{11}$ (e.g., Refs.~\cite{Anisovich:2009zy,Arndt:2006bf,Drechsel:2007if}). 
In the less known fourth resonance region, the two well-established $\Delta$~excitations 
$\Delta(1950)F_{37}$ and $\Delta(1920)P_{33}$ have been found to contribute 
(e.g., Ref.~\cite{Anisovich:2009zy}). The inclusion of polarization observables as additional 
constraints in the analysis of $\pi^0$~photoproduction data will not only help reveal 
contributions of resonances that couple only weakly to the $\pi^0$, but will also help to 
better understand the properties of these well-established resonances (e.g., the 
structure of the transition current).

\begin{figure*}[ht]
  \epsfig{file=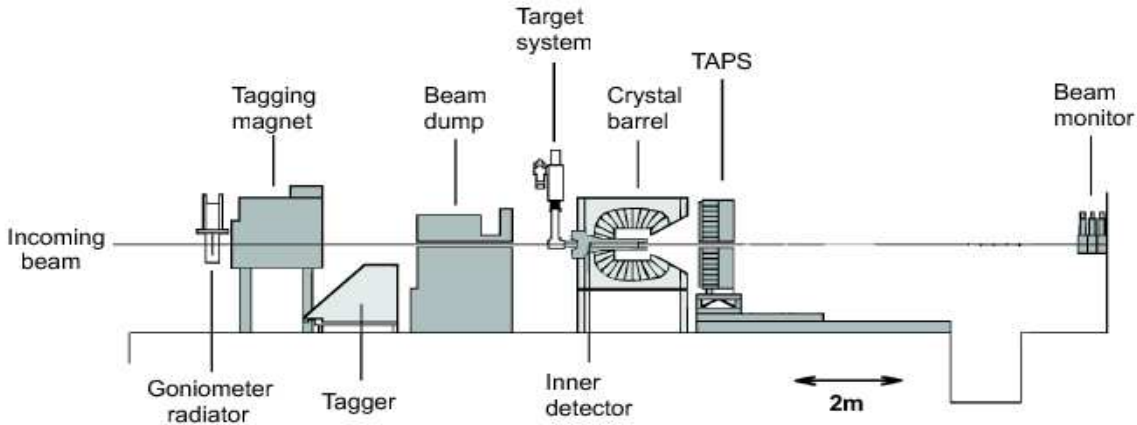,width=0.9\textwidth,height=0.38\textwidth}
  \vspace{-0.5cm}
  \caption{\label{Figure:Experiment} Experimental setup of CBELSA/TAPS in Bonn. 
     The electron beam delivered by the accelerator ELSA enters from the left side
     and hits the diamond crystal of the goniometer in front of the tagger magnet.}
\end{figure*}

In this paper, we present the beam asymmetry $\Sigma$ for the reaction:
\begin{align}
\vec{\gamma} p\to p\pi^0,&\quad\text{ where }\pi^0\to 2\gamma.\label{Reaction}
\end{align}
\noindent
The polarization data cover an incoming photon energy range between about 920 and 
1680~MeV and, in addition to $115^\circ < \theta_{\rm c.m.} < 155^\circ$, the most forward angular 
range of the $\pi^0$~meson $\theta_{\rm c.m.} < 60^\circ$.

The paper has the following structure. Section~\ref{Section:PreviousResults}
summarizes the data that were published before this analysis. An introduction to 
the CBELSA/TAPS experimental setup is given in Sec.~\ref{Section:ExperimentalSetup}.
The data reconstruction and selection are discussed in Sec.~\ref{Section:DataAnalysis},
and the extraction of beam asymmetries is described in Sec.~\ref{Section:ExtractionSigma}. 
Experimental results are finally presented in Sec.~\ref{Section:Results}.

\section{\label{Section:PreviousResults}Previous Results}
Cross section data for $\pi^0$~photoproduction were obtained and studied at many 
different laboratories over a wide kinematic range~\cite{Yoshioka:1980vu,
Bergstrom:1997jc,Beck:1990da,Beck:1997ew,Fuchs:1996ja,Krusche:1999tv,Schmidt:2001vg,
Blanpied:2001ae,Ahrens:2002gu,Bartholomy:2004uz,Bartalini:2005wx,vanPee:2007tw,Dugger:2007bt,Sumihama:2007qa}. A 
review of the main data sets and a corresponding comparison of their coverage in energy 
and solid angle can be found in Ref.~\cite{vanPee:2007tw}. 
%

Polarization observables for single-$\pi^0$ photoproduction have been determined mostly by
using a linearly polarized beam~\cite{Knies:1974zx,Alspector:1972pw,Bussey:1975mt,Bussey:1979wt,
Belyaev:1983xf,Blanpied:1992nn,Adamian:2000yi,Blanpied:2001ae,Schmidt:2001vg,Bartalini:2005wx,
Sumihama:2007qa,Elsner:2008sn}. In the following, a summary is given of the experiments 
performed after 1970 that allowed the extraction of the beam asymmetry~$\Sigma$. Most of 
the experiments accumulated data at very low energies ($< 500$~MeV); only very recently 
have data been taken above 1~GeV in the incoming photon energy.

In the 1970s, one of the earlier experiments used linearly polarized photons with 
energies from 610 to 940~MeV. The experiments were carried out by using the backscattered 
laser beam and the 82-in. bubble chamber at SLAC~\cite{Knies:1974zx}. 
At the Cambridge Electron Accelerator, beam asymmetries and cross sections at $\theta_{\rm c.m.} = 90^\circ$ were 
measured with photon energies that ranged from 0.8 to 2.2~GeV~\cite{Alspector:1972pw}. Finally, the 
Daresbury synchrotron allowed a study of the photon asymmetry over a range of photon 
energies from 1.2 to 2.8~GeV and over a range of $-t$ from 0.13 to 
1.4~(GeV/$c$)$^2$~\cite{Bussey:1975mt,Bussey:1979wt}.

Belyaev {\it et al.} measured $\Sigma$ in addition to the target asymmetry $T$ and
the double-polarization observable~$P$ by using linearly polarized photons and a 
transversely polarized proton target. The measurements were made in the energy range 
$E_\gamma\in [280,450]$~MeV and at $\pi^0$~c.m. angles between $60^\circ$ and $135^\circ$
\cite{Belyaev:1983xf}. 

Beck {\it et al.} measured differential cross sections at the electron accelerator 
MAMI (Mainz Microtron) between the threshold at 144~MeV up to photon energies of 157~MeV~\cite{Beck:1990da} 
as well as for energies between 270 and 420~MeV~\cite{Beck:1997ew}. Both experiments
used a linearly polarized photon beam produced via coherent bremsstrahlung. In 
Ref.~\cite{Beck:1997ew}, $\pi^0$~photoproduction was studied with the DAPHNE detector,
which covered $\sim 94\,\%$ of the solid angle.

In another experiment at MAMI, Schmidt {\it et al.} measured the 
photon asymmetry between threshold and 166~MeV by using the photon spectrometer TAPS. Total 
and differential cross sections were extracted simultaneously and were compared to predictions 
of chiral perturbation theory and low-energy theorems~\cite{Schmidt:2001vg}.

Blanpied {\it et al.} extracted unpolarized differential cross sections and 
beam asymmetry angular distributions at BNL by using LEGS for photon beam energies in 
the range from 213 to 333~MeV~\cite{Blanpied:1992nn,Blanpied:2001ae}. Final-state 
particles were detected in an array of six NaI crystals.

The Erevan group published data from several experiments. More recently, Adamian {\it et 
al.} extracted asymmetry data in the energy range 500--1000~MeV and for $\pi^0$~angles 
between $85^\circ$ and $125^\circ$ with energy and angle steps of 25~MeV and $5^\circ$, 
respectively~\cite{Adamian:2000yi}.
%

More recently, the GRAAL collaboration at ESRF in Grenoble extracted $\Sigma$ over 
a wide angular range, although limited to ${\rm cos}\,\theta_{\rm c.m.} < 0.7$. The data 
cover incoming photon energies between 550 and 1475~MeV~\cite{Bartalini:2005wx}. 
At GRAAL, Compton backscattering of low-energy photons off ultrarelativistic electrons 
reached almost 100\,\% beam polarization at the Compton edge.

The LEPS collaboration at SPring-8 in Hyogo, Japan, measured beam asymmetries for higher 
photon energies between $E_\gamma = 1500$ and 2400~MeV and, for the first time, at $\pi^0$
backward angles $-1 < {\rm cos}\,\theta_{\rm c.m.} < -0.6$~\cite{Sumihama:2007qa}. 
Backward-Compton scattering was applied by using Ar-ion laser photons with a 351-nm wavelength.

Recent CBELSA/TAPS asymmetry data cover photon energies between 760 and 1400~MeV 
and an angular range mostly in the backward direction of the $\pi^0$~meson ($110^\circ < 
\theta_{\rm c.m.} < 160^\circ$). For most of the energy bins, additional data points can be found in 
the angular region $50^\circ < \theta_{\rm c.m.} < 60^\circ$; there are a few data points at about 
$\theta_{\rm c.m.} = 40^\circ$~\cite{Elsner:2008sn}.

\section{\label{Section:ExperimentalSetup}Experimental Setup}
The results presented here are partially based on a reanalysis of the data discussed 
in Ref.~\cite{Elsner:2008sn}. The experiment was carried out at the electron accelerator facility 
ELSA~\cite{Hillert:2006yb} at the University of Bonn, Germany, by using a combination 
of the Crystal Barrel~\cite{Aker:1992} and TAPS~\cite{Novotny:1991ht,Gabler:1994ay} 
calorimeters. A schematic of the experimental setup at the ELSA facility is 
shown in Fig.~\ref{Figure:Experiment}.
\begin{figure}[t]
  \epsfig{file=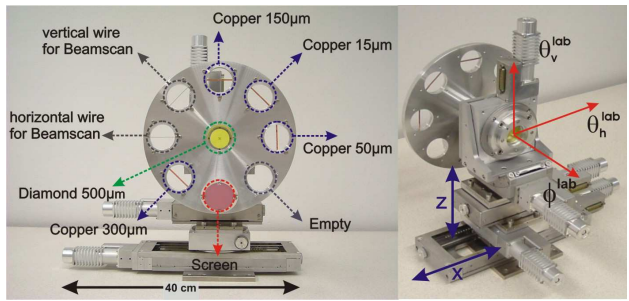,width=.46\textwidth}
    \caption{\label{Figure:Goniometer} (Color online) Photograph of the goniometer 
      setup for the CBELSA/TAPS experiment; picture taken from Ref.~\cite{Elsner:2008sn}.}
\end{figure}

Electrons with an energy of $E_0 = 3.175$~GeV were extracted from ELSA via slow (resonant) 
extraction. The electron beam then hit the radiator target positioned in front of the 
tagging magnet. The goniometer setup and its performance is fully described 
in Refs.~\cite{Elsner:2008sn,PhD:DanielElsner}. Since the development of the hardware and the 
production of linearly polarized photons is not part of the analysis presented here, only 
a very brief description of the setup is given. Several amorphous copper radiators with 
different radiation lengths surrounded the diamond crystal~(Fig.~\ref{Figure:Goniometer}). 
The crystal measured 500~$\mu$m in thickness and had a front surface of 4~x~4~mm;
it was glued to a $12.5~\mu$m kapton foil and was accurately positioned by a dedicated 
commercial five-axis goniometer. A wobble along the axes limited the maximum angular 
uncertainty to $\delta < 170~\mu$rad. All other uncertainties were negligible. 

The electrons undergoing the bremsstrahlung process were deflected in the dipole 
magnet according to their energy loss; the remaining energy was determined in a 
tagging detector consisting of 480~scintillating fibers above 14~scintillation 
counters in a configuration with adjacent paddles partially overlapping. The 
corresponding energy of an emitted photon was $E_\gamma = E_0 - E_{\rm e^-}$. 
Electrons not undergoing bremsstrahlung were deflected at small angles and were guided 
into a beam dump located behind the tagger detectors. The energy resolution is 
about 2~MeV for the high-energy photons and 25~MeV for the low-energy part of the 
bremsstrahlung spectrum.

For the energy calibration of the tagging system, a polynomial was determined in 
simulations using the measured field map of the bending magnet and the known positions 
of the fibers. The calibration was then cross-checked by measurements with the ELSA 
electron beam at two different energies. At 600 and 800~MeV, a low-current beam was 
guided directly into the tagger, while the magnetic field was slowly varied. More details
of the calibration can be found in Ref.~\cite{PhD:FrankKlein}.

\begin{figure}[t]
  \begin{center}
    \epsfig{file=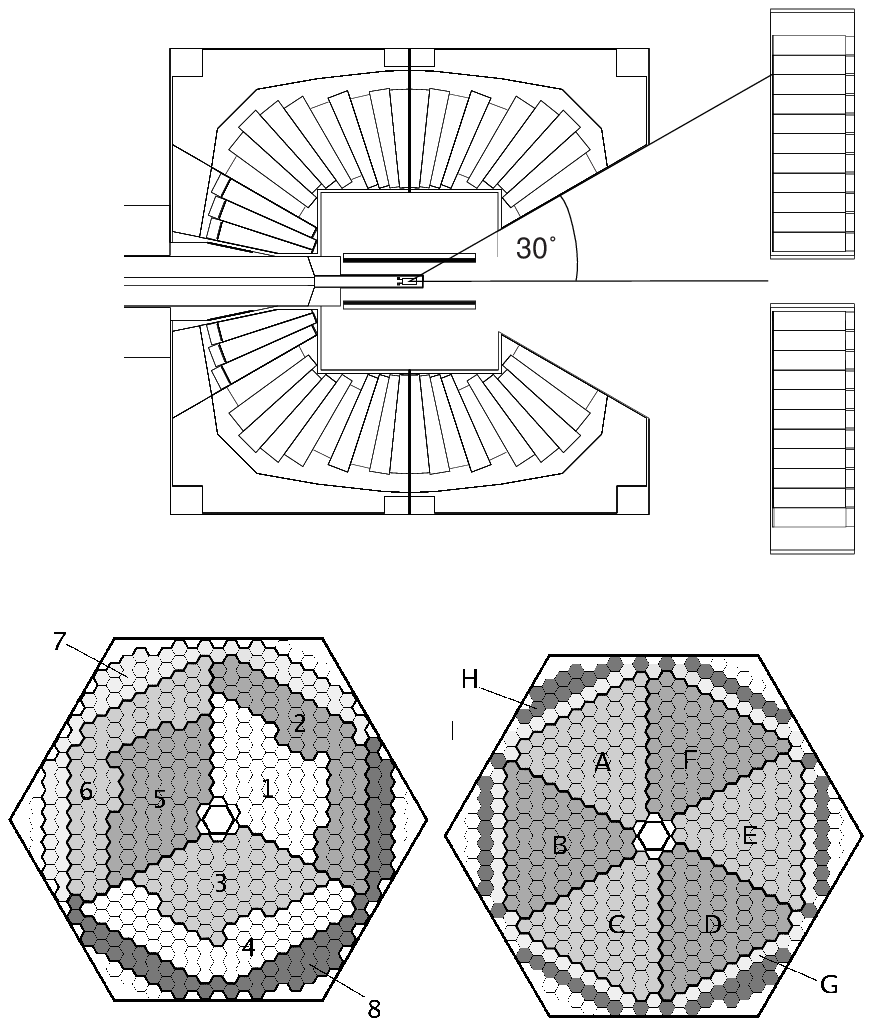,width=0.5\textwidth}
  \end{center}
  \caption{\label{Figure:CB-Luzy-H2} Top: Schematic of the liquid-hydrogen 
    target, scintillating-fiber detector, Crystal Barrel and TAPS calorimeters. 
    Bottom: Front view of TAPS; the left side shows the logical segmentation for the 
    LED-low trigger, the right side the logical segmentation for the LED-high trigger 
    (see text for more details).}
\end{figure}

\begin{figure*}[t]
  \epsfig{file=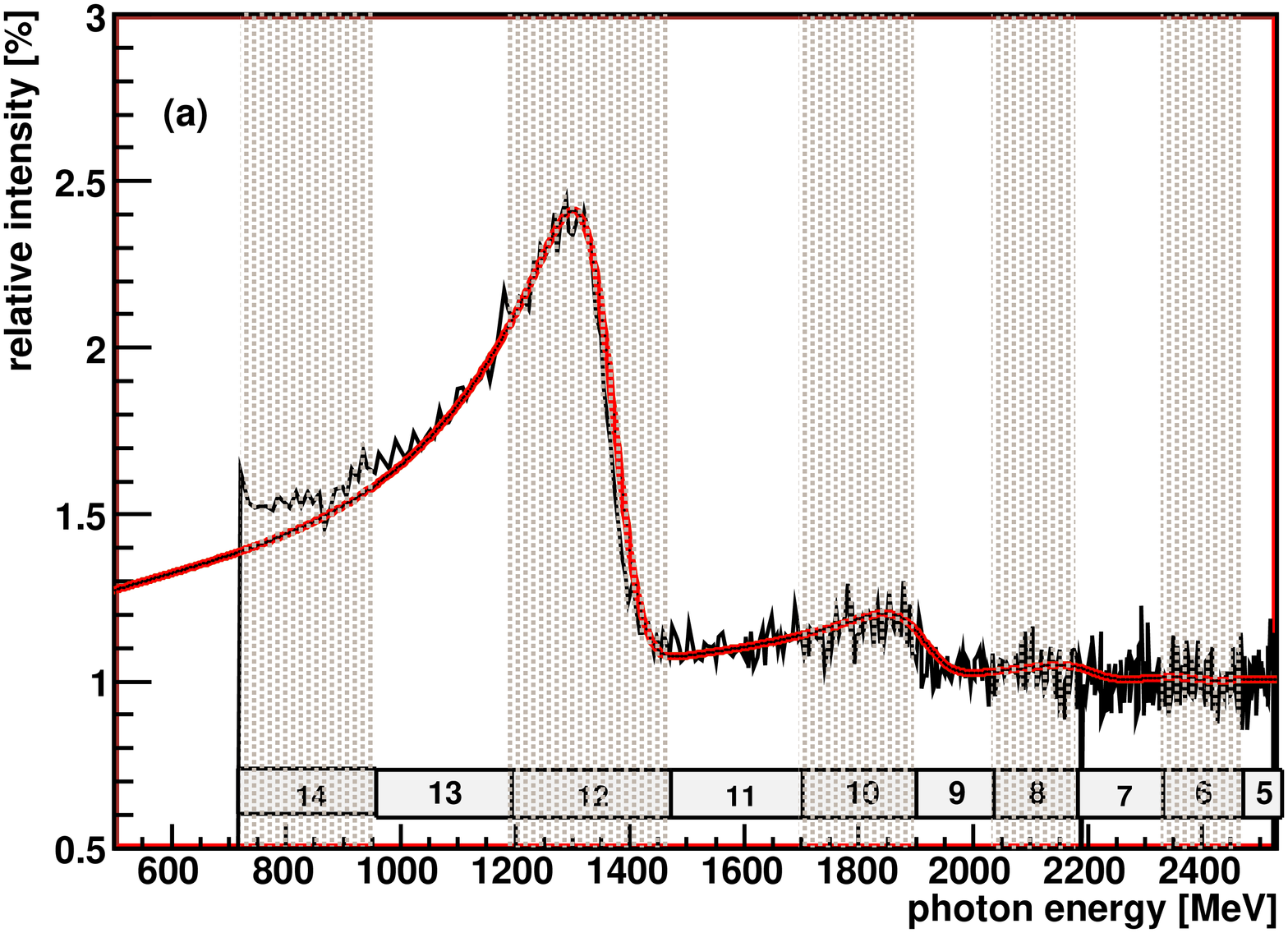,angle=-90,width=0.47\textwidth} \hfill
  \epsfig{file=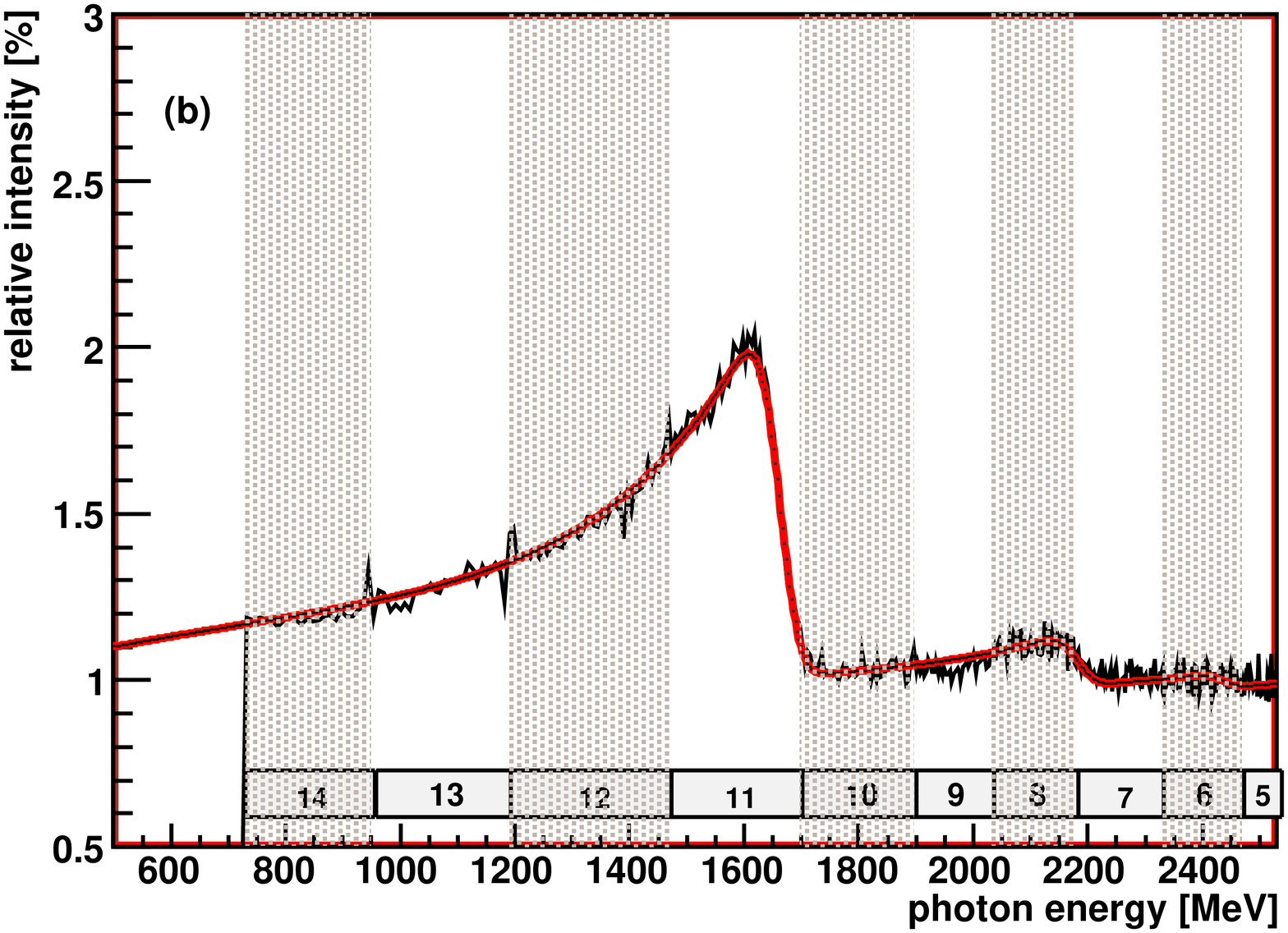,angle=-90,width=0.47\textwidth}
  \caption{\label{Figure:Intensities}(Color online) The measured coherent bremsstrahlung
    intensities normalized to an incoherent spectrum~\cite{Elsner:2008sn}. The full curve 
    shows how well the data are described by the model calculations~\cite{PhD:DanielElsner}. 
    For this experiment, the diamond radiator was oriented such that intensity maxima at 
    (a) $E_\gamma = 1305$~MeV and (b) $E_\gamma = 1610$~MeV were reached. The 
    boxes at the bottom of each distribution indicate the ranges covered by the 14 scintillation 
    counters of the tagger.}
\end{figure*}

The photons hit the liquid-hydrogen target in the center of the Crystal Barrel (CB) 
calorimeter. The target cell (5~cm in length, 3~cm in diameter) was surrounded by a 
scintillating-fiber detector~\cite{Suft:2005cq}, which provided an unambiguous impact 
point for charged particles (due to the arrangement of its three layers) leaving the 
target. The CB calorimeter in its CBELSA/TAPS configuration of 2002--2003 consisted 
of 1290 CsI(Tl) crystals with a length of 16~$X_R$. The modules have an excellent 
photon detection efficiency; a detailed description can be found in Ref.~\cite{Aker:1992}. 
For this series of experiments, the (downstream) rings 11--13 were removed to combine 
the detector with TAPS in the forward direction. The CB calorimeter covered the complete 
azimuthal angle and polar angles from $30^\circ$ to $168^\circ$. All crystals are of 
trapezoidal shape pointing to the center of the target~(Fig.~\ref{Figure:CB-Luzy-H2}, 
top).

The TAPS detector consisted of 528 hexagonal BaF$_2$ crystals with a length of about 
12~$X_R$. It was configured as a hexagonal wall serving as the forward end cap of 
the CB calorimeter (Fig.~\ref{Figure:CB-Luzy-H2}, bottom). TAPS provided 
a high granularity in the forward direction for polar angles between $5^\circ$ 
and $30^\circ$ (full $\phi$~coverage). A 5-mm thick plastic scintillator in front of 
each TAPS module allowed the identification of charged particles. The combination of 
the CB and TAPS calorimeters covered 99\,\% of the $4\pi$ solid angle and 
served as an excellent setup to detect multiphoton final states.

The fast response of the TAPS modules provided the first-level trigger. The second-level 
trigger was based on a cellular logic (FACE), which determined the number of clusters 
in the barrel. The trigger required either two hits above a low-energy threshold in TAPS 
(LED-low) or one hit above a higher-energy threshold in TAPS (LED-high) in combination 
with at least two FACE clusters. The shape of the logical segmentation for the TAPS 
trigger is shown in Fig.~\ref{Figure:CB-Luzy-H2} (bottom).

\subsection{\label{Subsection:Goniometer}Linearly polarized photons}
Two methods are usually applied for preparing a linearly polarized photon beam: coherent 
bremsstrahlung and Compton backscattering. In the latter technique, linearly polarized
laser photons are backscattered off a high-energy electron beam 
(e.g., Refs.~\cite{Nakano:2001xp,Bartalini:2005wx}). The degree of polarization that can be achieved by
using this technique is proportional to that of the initial laser beam. Although high degrees 
of polarization can, in principle, be reached, the photon beam intensities are usually lower 
than those from coherent bremsstrahlung because of limitations resulting from the operation 
of a multiuser storage ring. In contrast, many facilities have successfully produced 
linearly polarized beams by using coherent electron 
bremsstrahlung~\cite{Lohmann:1994vz,Elsner:2008sn}, where the recoil momentum of the
recoiling nucleus embedded in the crystal is transferred to the crystal lattice. For 
the CBELSA/TAPS experiment, a diamond crystal was used. For certain orientations of this 
diamond, the recoil momentum can be entirely transferred to the crystal; this defines 
the deflection plane of the electrons and results in a strong linear polarization of 
the photon beam.

%
For the beam asymmetry data presented in this paper, the crystal alignment was achieved 
by the so-called {\it Stonehenge Technique}~\cite{Livingston:2008hv}. An overview of the 
alignment process for the CBELSA/TAPS goniometer, which includes a brief description of the 
Stonehenge Technique, is given in Ref.~\cite{Elsner:2008sn}. The stability of the beam position 
was monitored online to preserve the alignment during the experiment. The coherent peak 
itself was used for this procedure because the position of the coherent edge in the energy 
spectrum is extremely sensitive to the angle of the incident beam~\cite{PhD:FrankKlein}. 

\begin{figure*}[t]
  \epsfig{file=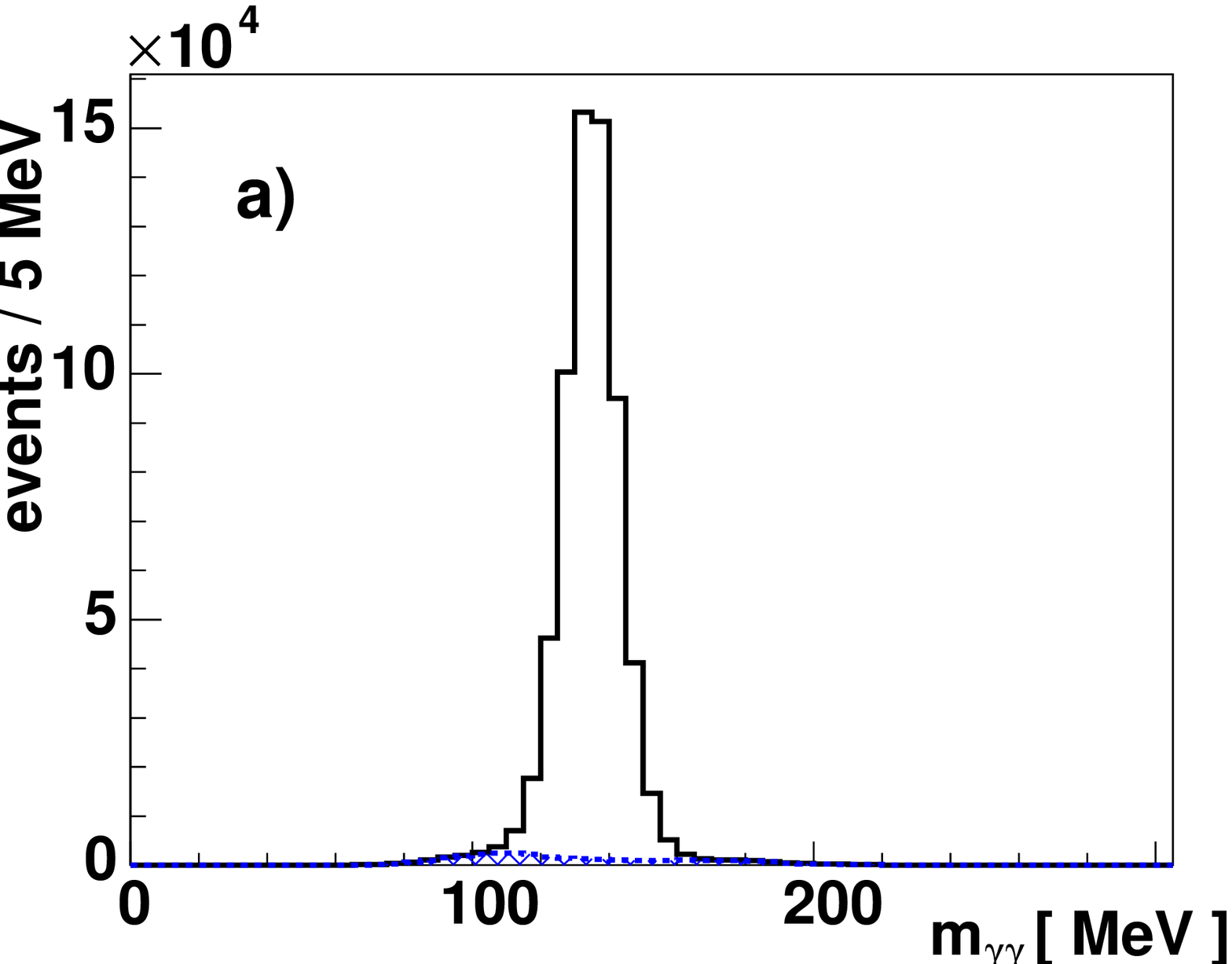,width=.32\textwidth} \hfill
  \epsfig{file=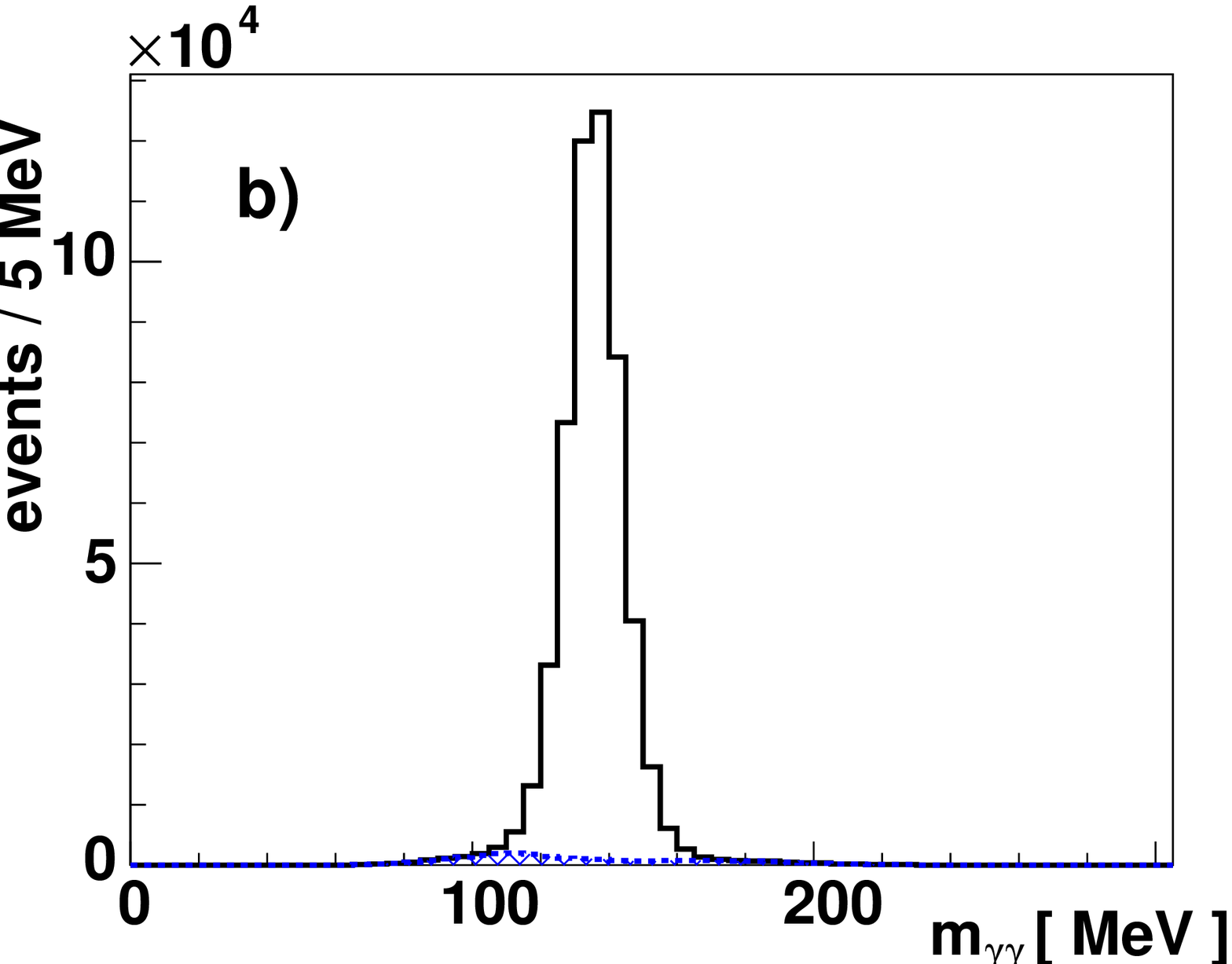,width=.32\textwidth} \hfill
  \epsfig{file=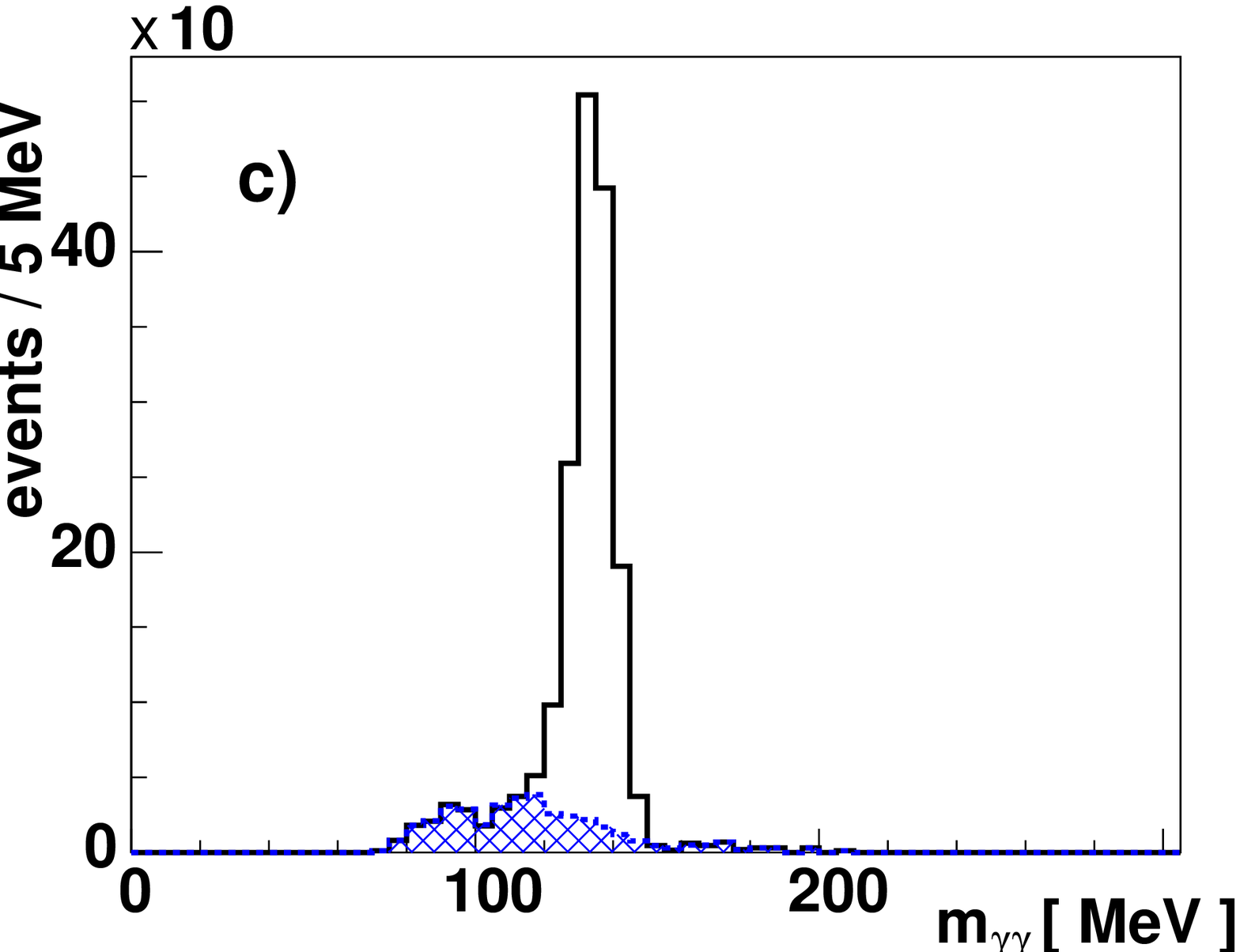,width=.32\textwidth}
  \caption{\label{Figure:Masses} (Color online) Invariant $\gamma\gamma$-mass spectra for 
    the reaction $\gamma p\to p\gamma\gamma$ using data with the coherent peak at 1305~MeV 
    (left) and  at 1610~MeV (center); confidence level cuts were applied at $10^{-2}$. The 
    $\pi^0$~mesons are observed with very little background. On the right, a mass spectrum 
    for a forward bin is shown at $E_\gamma = 1097$~MeV (bin width 33 MeV) and $\theta_{\rm c.m.} 
    = 25^\circ\pm 5^\circ$. The colored area (bottom) indicates the background (for the background 
    determination, see Sec.~\ref{Subsection:Background}).}
\end{figure*}

The degree of linear polarization was determined in Ref.~\cite{PhD:DanielElsner} by comparing the 
measured photon spectrum with a model calculation using the ANB (analytic bremsstrahlung calculation) 
software~\cite{Natter:}. Fig.~\ref{Figure:Intensities} shows photon intensity spectra normalized to incoherent 
spectra~\cite{Elsner:2008sn} for the two different positions of the coherent edge used in this analysis. The 
curves represent calculations by using an improved version of the original ANB code~\cite{PhD:DanielElsner}, 
which takes the effects of beam divergence, beam spot size, energy resolution, and multiple 
scattering into account. 
The description of the measured spectra is excellent at all energies and coherent peak positions. An absolute 
error of $\delta P_\gamma < 0.02$ is estimated by using variations of the calculated relative intensity by 
$\pm\,5\,\%$~\cite{Elsner:2008sn}. These worst-case estimates account for deviations from the shape of 
the spectrum caused by combined statistical and systematic effects.
%

\section{\label{Section:DataAnalysis}Preparation of Final State}
The data presented here were accumulated in March and May of 2003 in two run periods with 
an ELSA beam energy of 3.175~GeV. Events for coherent peak positions at 1305 and 
1610~MeV 
were recorded. 
These CBELSA/TAPS polarization data were used to extract the beam asymmetries for a large variety 
of photoproduction reactions~\cite{Elsner:2008sn,Elsner:2007hm,Gutz:2008zz,Gutz:2009zh,:2008gs}. 
The analysis discussed here, for the peak position at 1305~MeV, is partially a reanalysis of 
the data published in Ref.~\cite{Elsner:2008sn}. The event reconstruction and selection of the 
$\pi^0$~channel for Reaction~(\ref{Reaction}) is presented in this section. A total number of 
$\sim 1.06\times 10^6$~$\pi^0$ events has been included in this analysis.

\subsection{\label{Subsection:EventReconstruction}Event reconstruction}
Events with two or three (neutral or charged) particles in the final state were selected. The
experimental setup allows the identification of charged clusters in TAPS by using the plastic 
scintillators mounted in front of each BaF$_2$ crystal. The efficiencies of these (photon)-veto
detectors were determined and were modeled in the Monte Carlo program. Although these detectors have 
been used in a recent extraction of unpolarized $\eta$~and $\eta\,^\prime$ differential cross 
sections~\cite{Crede:2009zz}, we decided not to employ this information in the analysis
to avoid a possible $\phi$~dependence of the data on these detector components. Instead, the
proton in all events with three particles was identified by successively assigning the proton tag 
to each final-state particle (and by assuming the remaining two particles are photons) and then by 
testing the hypothesis $\gamma p\to p\gamma\gamma$ in a 1C~kinematic fit only requiring energy 
and momentum conservations. Simultaneously, all possible tagger photons were tested. A prompt 
coincidence within $-5$ to $+15$~ns between a particle in TAPS and an electron in the tagger was required
to reduce time-accidental background. The best fit based 
on its $\chi^2$~probability or confidence level (CL) defined the proton as well as the initial photon 
and its corresponding energy.

On average, proton clusters in the calorimeters are much smaller than photon clusters and
can sometimes consist of only one or two crystals; this provides an insufficient resolution. For
this reason, proton identification was used only to remove the proton from the list of
final-state particles. The proton momentum was then reconstructed from the event kinematics
in ``missing-proton'' kinematic fitting.

The use of kinematic fitting in CBELSA/TAPS data analyses has been described in more detail 
in Ref.~\cite{Crede:2009zz}. In this analysis, all events were subject to the hypothesis:
\begin{eqnarray}
\label{Equation:Hypothesis}
  \gamma p\,\to\, p\,n_\gamma \gamma\,,
\end{eqnarray}
which simply imposes energy and momentum conservations without a $\pi^0$ mass constraint.
Figure~\ref{Figure:Masses} shows the remaining invariant $\gamma\gamma$~mass for all events 
satisfying Eq.~(\ref{Equation:Hypothesis}) at a CL~$> 10^{-2}$. A clear
peak for the $\pi^0$~meson is visible. The background underneath the peak was subtracted 
for every $(E_\gamma,\,\theta_{\rm c.m.},\,\phi$)~bin by using the so-called $Q$-factor method 
described in the following section.

\subsection{\label{Subsection:Background}Background subtraction}
Mass distributions for ($E_\gamma,\theta_{\rm c.m.},\phi$)~bins in the forward direction of the
$\pi^0$~meson show some residual background under the meson peak. The separation of background events
from signal events is typically done by using the side-band subtraction method. In this approach,
events from outside the signal region are subtracted from those inside the signal region to
remove the background from the distribution. 

We decided to use an event-based approach called the $Q$-factor method, which assigns a signal probability $Q_i$ to each 
event. The approach is described in detail in Ref.~\cite{Williams:2008sh}. In most of our forward
bins, the functional form of the background shape $B(m,\vec{\xi})$ is unknown, where each 
$\gamma p\to p\gamma\gamma$~event has kinematic variables $\vec{\xi} = (E_\gamma,\,{\rm cos}
\,\theta_{\rm c.m.}^{\gamma\gamma},\,\phi_{\gamma\gamma},\,M_{\gamma\gamma},\,\,\phi^\ast,\,\theta^\ast)$. The two variables 
$\phi^\ast$ and $\theta^\ast$ denote the azimuthal and polar angles in the rest frame of the two photons; these angles 
are measured with respect to the coordinate system formed by the unit vector $\hat{z}^{\prime}$ in the direction of the 
two-photon system in the c.m. frame, the unit normal $\hat{y}^{\prime}$ to the reaction plane (defined below), and 
$\hat{x}^{\prime} = \hat{y}^{\prime} \times \hat{z}^{\prime}$. The azimuthal angle $\phi^\ast$ is also given by the angle 
between the reaction plane and the two-photon decay plane, where the reaction plane is spanned by the beam axis and the 
recoiling proton (or the two-photon system); both of these planes are formed by particles in the c.m. system. 
The invariant $\gamma\gamma$~mass was chosen as the reference variable for which the background dependence was studied. 
The distance between any two events in the space spanned by~$\vec{\xi}$ is given by~\cite{Williams:2008sh}
\begin{eqnarray}
\label{Equation:Distance}
  d^2_{ij}\,=\,\sum^6_{k=1}\,\biggr[\,\frac{\xi^i_k - \xi^j_k}{r_k}\,\biggl]^2\,, 
\end{eqnarray}
\noindent
where $r_k$ denotes the range of $\xi_{k}$ and the reference variable is excluded in the sum. We 
then found the closest 100 events for each event $i$, with kinematics $\vec{\xi}_i$ and $\gamma\gamma$~mass 
$M_i$, according to Eq.~(\ref{Equation:Distance}). Since these 100 events occupy a very small 
region around $\vec{\xi}_i$, a linear approximation is validated for the mass dependence of 
the background. A Gaussian line shape was used to model the $\pi^0$~signal. We have used the 
unbinned maximum likelihood method to obtain the parameters describing the mass distributions. 
By using these fit results, the expected number of signal and background events, denoted as $s_i$ 
and $b_i$, respectively, can be calculated at $M_i$ and for each event, the $Q$-factor can be 
written as
\begin{eqnarray}
\label{Equation:Q}
  Q_i\,=\,\frac{s_i}{s_i + b_i},\quad {\rm where}~~N_{\rm signal}\,=\,\sum_i^N\,Q_i\,.
\end{eqnarray}
\noindent
This method delivered a reliable subtraction of the background from our mass distributions. 
The background visible in Fig.~\ref{Figure:Masses} has been determined using this method.
The $Q$-factor errors (or systematic uncertainties on signal-yield extractions) contribute
strongly to the total systematic uncertainty of the extracted polarization observables. A full 
discussion of the error estimation and event correlations goes beyond the scope of this paper 
and can be found in Ref.~\cite{Williams:2008sh}.

\subsection{\label{Section:MonteCarloSimulations}Monte Carlo simulations}
The performance of the detector was simulated in GEANT3-based Monte Carlo studies. We used 
a program package that was built upon a program developed for the CB-ELSA experiment. 
The Monte Carlo program accurately reproduces the response of the TAPS and CB 
crystals when hit by a photon.

The acceptance for Reaction~(\ref{Reaction}) was determined by simulating events that were 
evenly distributed over the available phase space. The Monte Carlo events were analyzed by using 
the exact same reconstruction criteria as the (real) measured data. 
The same 1C hypothesis was tested in the kinematic fits, and events were selected with the same 
CL cuts. The acceptance is defined as the ratio of the number of generated to 
reconstructed Monte Carlo events:
\begin{equation}
  A_{\gamma p\to p\,X}=\frac{N_{\rm rec,MC}}{N_{\rm gen,MC}}
     \qquad (X=\pi^0\,)\,.
\end{equation}
\noindent
In the analysis presented here, we have required the acceptance to be at least 8\,\% in ($E_\gamma,
\theta_{\rm c.m.}\,$) bins and removed the other data points from the analysis. 
\begin{figure}[t]
  \epsfig{file=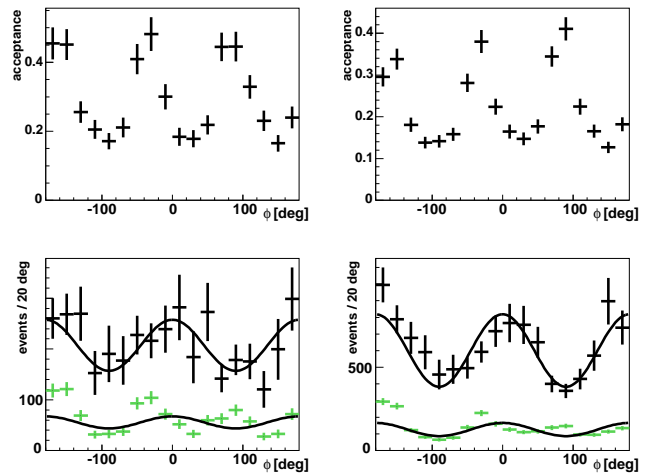,width=.49\textwidth}
  \vspace{-5mm}
  \caption{\label{Figure:TriggerPhi} (Color online) Influence of the hardware trigger. The 
    top row shows the acceptance for $E_\gamma\sim 1560$~MeV as well as $\theta_{\rm c.m.} = 
    15^\circ\pm 5^\circ$ (left) and $\theta_{\rm c.m.} = 25^\circ\pm 5^\circ$ (right), respectively. The three-peak 
    structure caused by the boundaries between the trigger segments is visible (see text for more
    details). The corresponding data $\phi$~distributions are given in the bottom row. The 
    colored distributions (bottom) show the uncorrected distributions and the black data points (top) 
    show the acceptance-corrected data. The improvement can clearly be observed.}
\end{figure}

\begin{figure*}[t]
    \epsfig{file=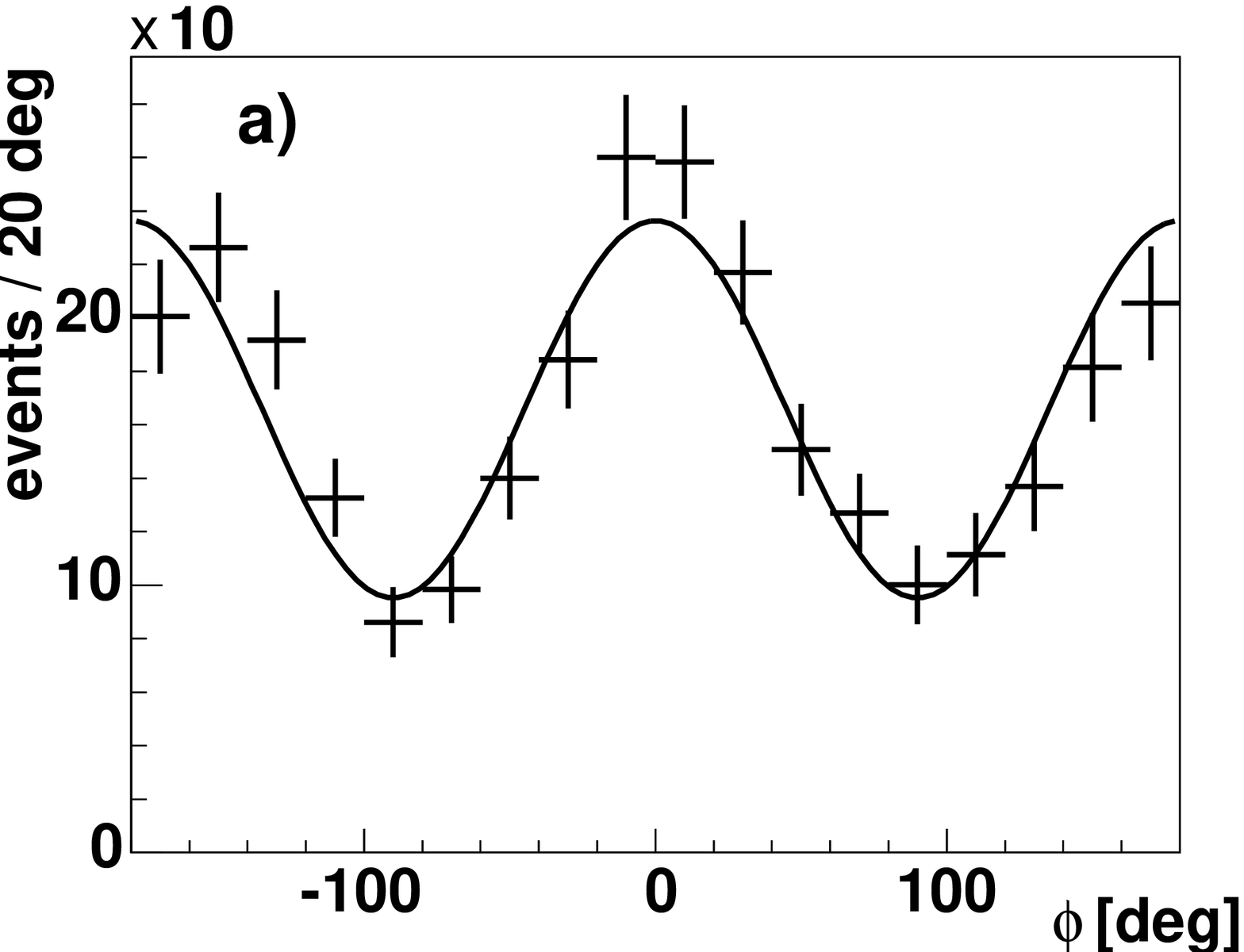,width=.32\textwidth} \hfill
    \epsfig{file=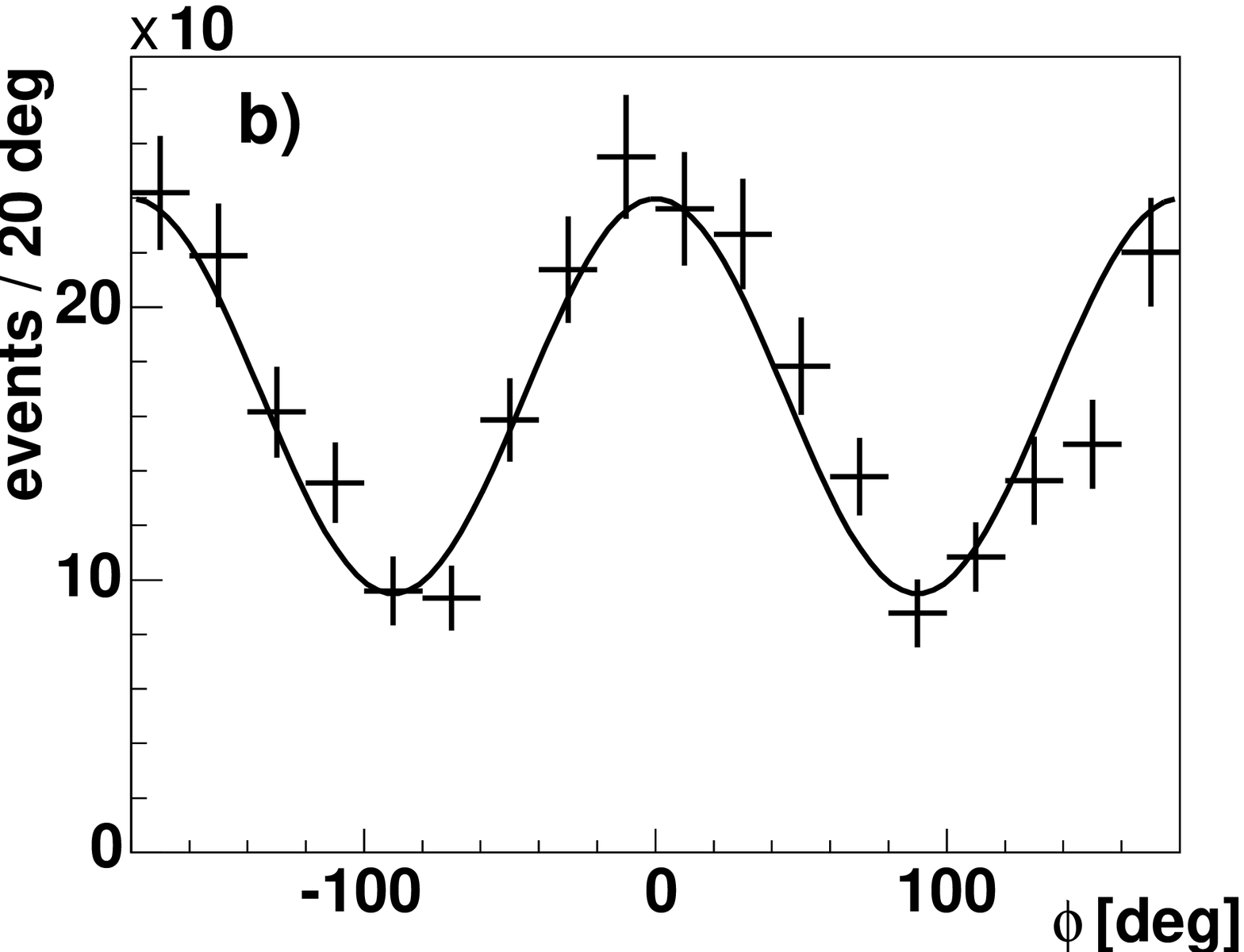,width=.32\textwidth} \hfill
    \epsfig{file=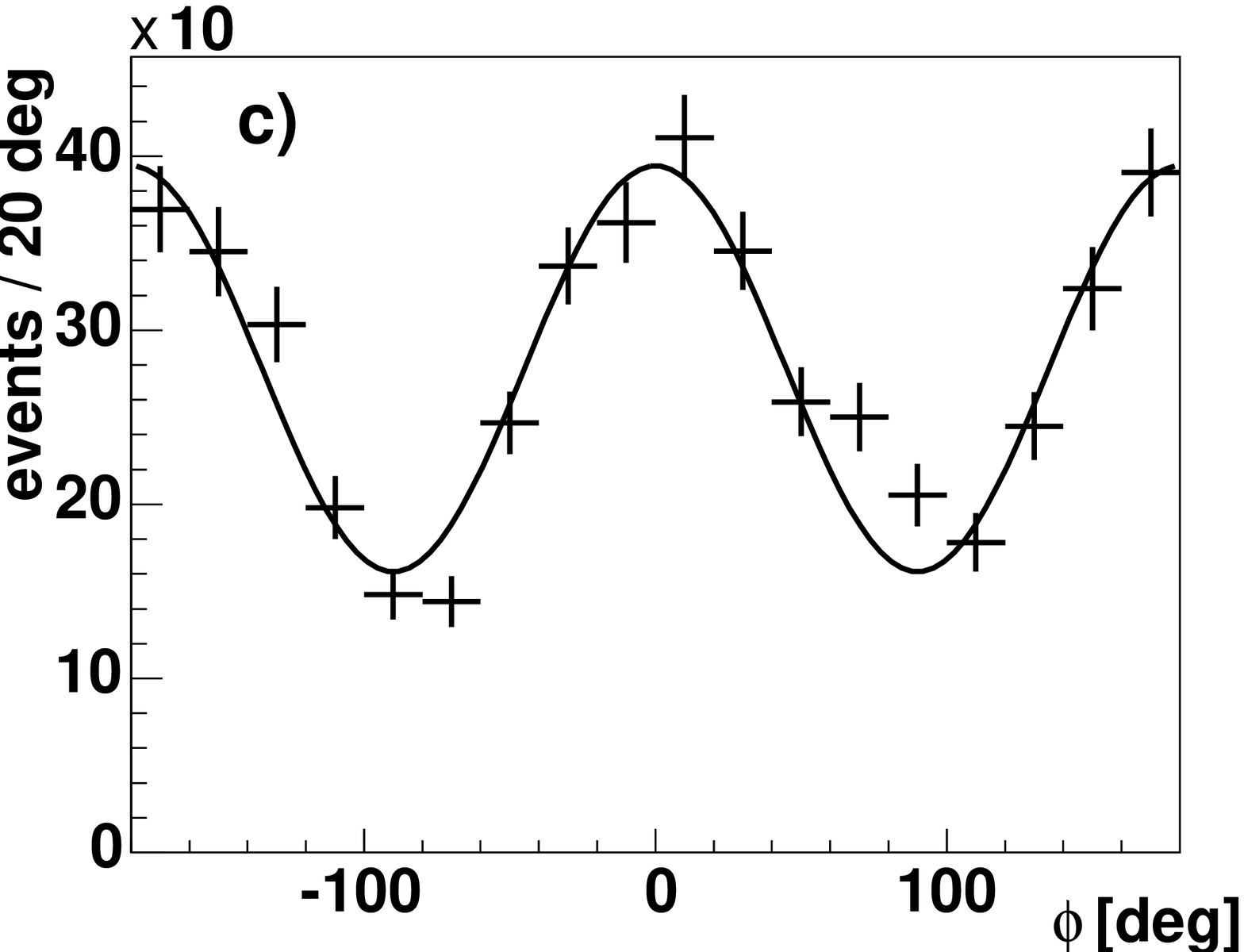,width=.32\textwidth} \hfill
    \caption{\label{Figure:Phi}Typical $\phi$~distributions for forward-angle bins at 
       $\theta_{\rm c.m.} = 35^\circ$ and $E_\gamma$ = 1229~MeV (left),~1295~MeV (middle), and
      1625~MeV (right). We have chosen 18 bins in the azimuthal angle $\phi$, which corresponds 
       to a bin width of $20^\circ$.}
\end{figure*}

For the extraction of beam asymmetries, it is important to study possible systematic 
(nonphysics related) contributions to the $\phi$~distributions. Of particular importance 
is the influence of the hardware trigger. It required either a hit above a lower-energy 
threshold in at least two different segments of the TAPS LED-low logical segmentation 
(Fig.~\ref{Figure:CB-Luzy-H2}, bottom left), or a higher-energy hit in one of 
the TAPS high-trigger segments (Fig.~\ref{Figure:CB-Luzy-H2}, bottom right) in combination 
with at least two clusters in the CB  was needed (trigger condition 2). If the 
event kinematics is such that only one particle hits TAPS (possibly leading to condition 2), 
condition 1 can also be fulfilled simultaneously when the hit occurs close to the edge of 
a segment. The electromagnetic shower leaking into the adjacent trigger segment then increases 
the trigger efficiency along the boundaries, imposing a modulation in the $\phi$~distribution. 
Figure~\ref{Figure:TriggerPhi} shows examples of this effect for $E_\gamma\sim 1560$~MeV. The 
three peaks are caused by the three boundaries in the logical segmentation of the LED-low trigger. 
Since this effect is $\phi$~dependent, it can, in some cases, which depend on event kinematics, 
significantly contribute to the $\phi$~modulations. The $\phi$~distributions in the forward 
region have, thus, been acceptance corrected to account for the described trigger effect. 

\section{\label{Section:ExtractionSigma}Extraction of $\Sigma$}
%

\begin{figure}[b]
  \epsfig{file=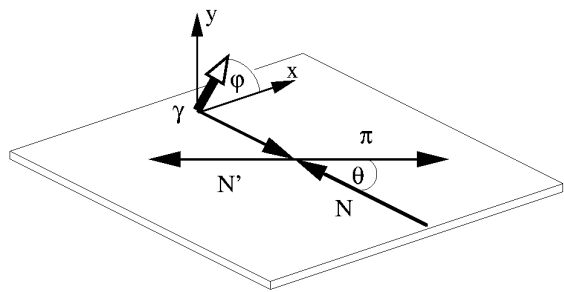,angle=-90,width=.47\textwidth}
  \vspace{-0.5cm}
  \caption{\label{Figure:Reaction}Sketch of the $\gamma p\to p\pi^0$ reaction 
    in the c.m. system; the open (white) arrow indicates the 
    linearly polarized photon.}
\end{figure}

The polarized cross section in single-$\pi$~photoproduction with linearly polarized
photons is proportional to the unpolarized cross section $(d\sigma/d\Omega)_0$ and is 
given by
\begin{eqnarray}
  \frac{d\sigma}{d\Omega}\,=\,\biggr(\frac{d\sigma}{d\Omega}\biggl)_0\,(\,1 - P_l\,\Sigma\,{\rm cos}
  \,(2\varphi)\,)\,,
\end{eqnarray}
where $P_l$ denotes the degree of linear beam polarization at an angle $\varphi$ with
respect to the reaction plane, which is spanned by the incoming photon and the recoiling
nucleon in the c.m. system. The reaction is schematically shown in Fig.~\ref{Figure:Reaction}. In the
experiment, the orientation of the photon polarization is given in the laboratory frame
by an angle $\alpha$ and, thus, $\varphi = \alpha - \phi$. For our measurements,
the diamond crystal was oriented such that the direction of the beam polarization was
perpendicular to the floor of the experimental area ($\alpha = \pi/2$):
\begin{eqnarray}
  \frac{d\sigma}{d\Omega}\,=\,\biggr(\frac{d\sigma}{d\Omega}\biggl)_0\,(\,1 + P_l\,\Sigma\,{\rm cos}
  \,(2\phi)\,)\,.
\end{eqnarray}
If the detector setup is invariant with respect to the azimuthal angle, then the 
observable $\Sigma$ can be extracted as the amplitude of the $\phi$~modulation of the
$\pi^0$~meson corrected for the degree of polarization.

Figure~\ref{Figure:Phi} shows typical $\phi$ distributions in the forward region. From 
fits to these azimuthal event distributions using a function of the form
\begin{eqnarray}
\label{Equation:Fit}
  f(\phi)\,=\,A\,+\,B\,{\rm cos}\,(2\phi)\,,
\end{eqnarray}
the product of beam asymmetry and photon polarization $P_l\,\Sigma$ is given by the
ratio $B/A$ for each bin of photon energy and $\pi^0$ polar angle $\theta_{\rm c.m.}$. 

\begin{figure*}[t]
  \epsfig{file=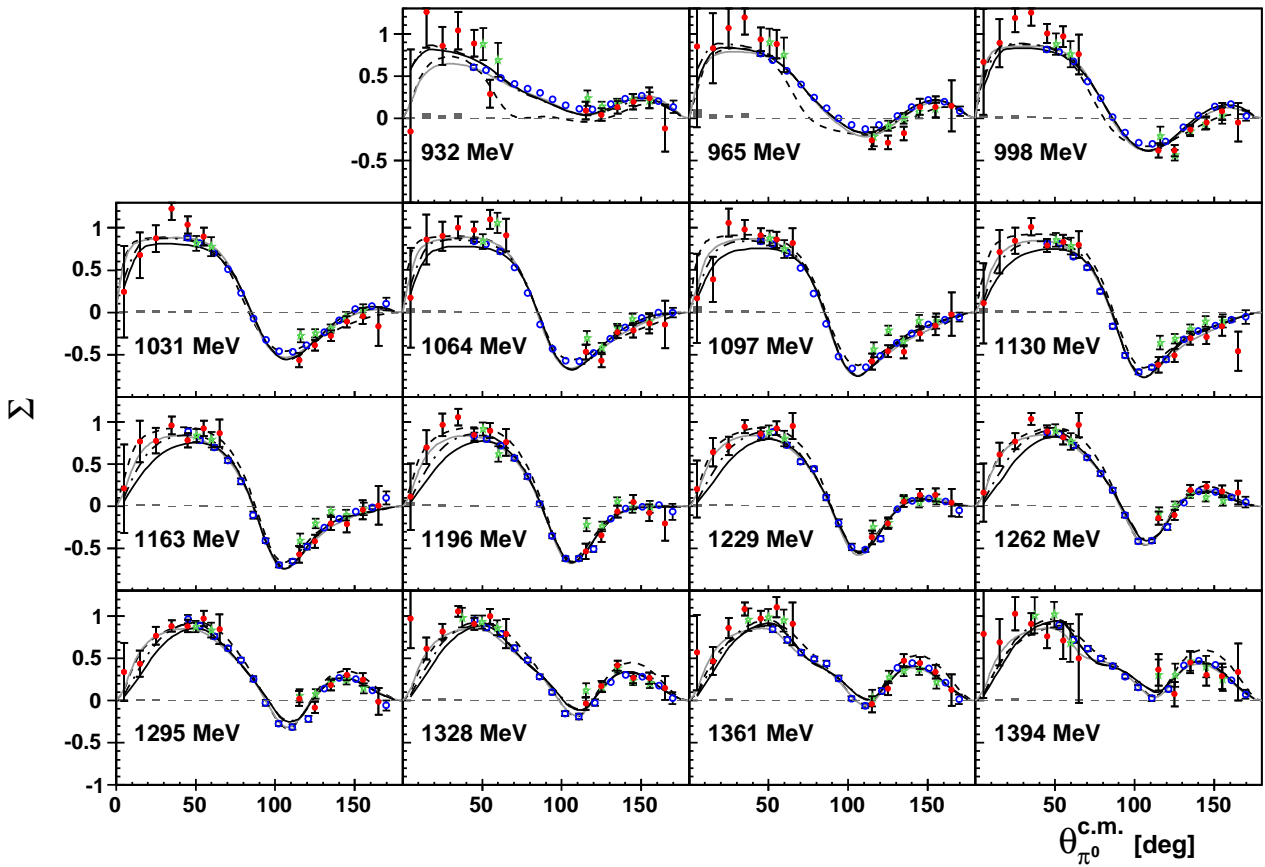,width=1.0\textwidth}
  \caption{\label{Figure:Results1350}(Color online) The photon beam asymmetries extracted 
      from the data set with a coherent peak position at 1305~MeV. The filled (red) circles
      {\color{red} ($\bullet$)} denote this analysis, the (green) stars {\color{green} 
      ($\ast$)} denote our previous CBELSA/TAPS analysis~\cite{Elsner:2008sn}, and the open
      (blue) circles {\color{blue} ($\circ$)} denote the GRAAL results~\cite{Bartalini:2005wx}. 
      The black solid line shows the recently published solution of the Bonn-Gatchina partial 
      wave analysis (PWA)~\cite{Anisovich:2009zy}, the gray solid line denotes the SAID SP09
      prediction~\cite{said:database,Dugger:2009pn}, the dashed black line shows the 
      recent MAID solution~\cite{Drechsel:2007if}, and the dashed-dotted line shows a new 
      preliminary solution of the Bonn-Gatchina PWA that includes the results of this analysis
      \cite{Andrey:private}. The width of the energy bins is 33~MeV, consistent with the earlier 
      published results. The energy of the bin centers is given in each distribution.}
\end{figure*}

\subsection{Systematic uncertainties\label{Subsection:SystematicUncertainties}}
\begin{figure*}[t]
  \epsfig{file=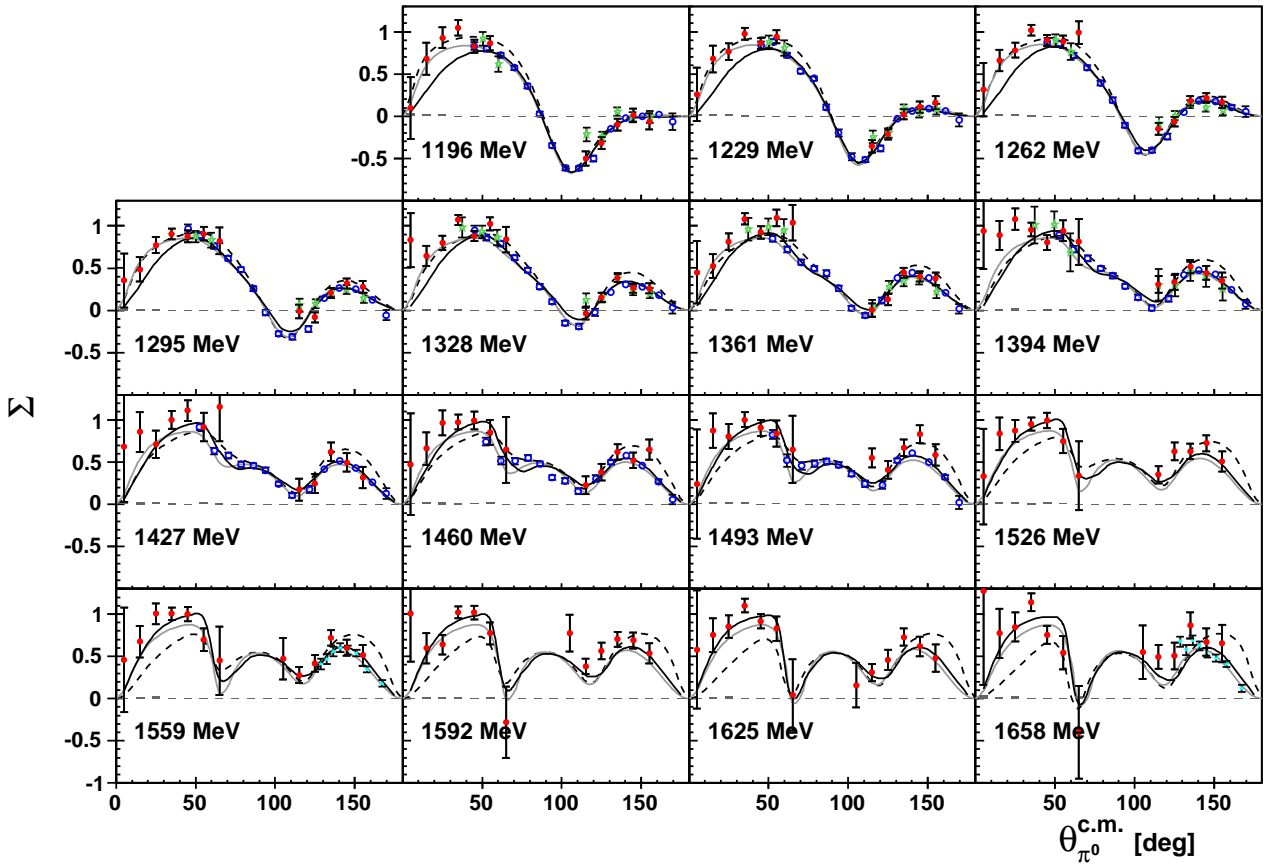,width=1.0\textwidth}
  \caption{\label{Figure:Results1600}(Color online) The photon beam asymmetries extracted 
      from the data set with a coherent peak position at 1610~MeV. The filled (red) circles
      {\color{red} ($\bullet$)} denote this analysis, the (green) stars {\color{green} 
      ($\ast$)} denote our previous CBELSA/TAPS analysis~\cite{Elsner:2008sn}, the open (blue) 
      circles {\color{blue} ($\circ$)} denote the GRAAL results~\cite{Bartalini:2005wx}, and the 
      (blue-green) stars {\color{cyan} ($\ast$)} above 1500~MeV denote recent LEPS results~\cite{Sumihama:2007qa}. 
      The black solid line shows the recent solution of the Bonn-Gatchina 
      PWA~\cite{Anisovich:2009zy}, the gray solid line denotes the SAID SP09
      prediction~\cite{said:database,Dugger:2009pn}, and the dashed black line shows the
      recent MAID solution~\cite{Drechsel:2007if}. The width of the energy bins is 33~MeV, consistent 
      with the earlier published results. The energy of the bin centers is given in each 
      distribution. For energies below 1400~MeV, we have averaged the results from both data 
      samples.}
\end{figure*}

The detection efficiency usually has a weak influence on polarization observables.
Most acceptance effects will drop out in the ratio $B/A$~[Eq.~(\ref{Equation:Fit})] if the 
bin sizes are small compared to the variation of the acceptance. 
Since the extraction of beam asymmetries is based on fits to $\phi$~distributions, the statistical and systematic 
errors of~$\Sigma$ cannot easily be separated. For this reason, the error bars (of the data points) 
in Figs.~\ref{Figure:Results1350} and~\ref{Figure:Results1600} consist of both contributions;
these errors also include the upper limit on the error of the degree of polarization $\delta 
P_\gamma = 0.02$. Further systematic uncertainties are given separately and are added to the 
error band. 

Systematic errors due to the background subtraction scheme were estimated by performing the following procedure.
The beam asymmetries were extracted by using $\phi$~distributions that have statistical errors determined from the
number of events in each bin. In a separate analysis step, the beam asymmetries were found by using $\phi$~distributions in
which the error of each bin is the quadratic sum of the statistical error and the error from the background
subtraction. The difference between these results gave an estimate of the systematic error for each data point.    

Further contributions to the systematic uncertainties were determined from Monte Carlo studies and
acceptance corrections. The reconstruction of neutral pions and the identification of final 
states required a sequence of cuts that included the use of kinematic fitting (CL cuts). 
Error contributions that account for the slightly different 
effects of CL cuts on data and Monte Carlo events (for the acceptance-corrected 
forward bins) are estimated at the 3\,\%~level, and they are included in the remaining systematic error 
plotted along the $\Sigma=0$~line in each distribution of Figs.~\ref{Figure:Results1350} 
and~\ref{Figure:Results1600}.

Additional sources of systematic errors are uncertainties with regard to possible unknown 
fluctuations of electronic equipment that contribute to the $\phi$~modulations and a possible 
offset of the photon beam. While electronic fluctuations have not been studied further,
the beam offset was assumed to be shifted by less than 2~mm off axis at the target position. 
A contribution of such a small offset to the beam asymmetry was found to be negligible.

%

\section{Experimental Results \label{Section:Results}}
Figure~\ref{Figure:Results1350} shows the $\pi^0$ beam asymmetries for our data set with a 
coherent peak position at 1305~MeV. The unusual energy bin width of 33~MeV was chosen to 
facilitate the comparison with the GRAAL~\cite{Bartalini:2005wx} and the previous CBELSA/TAPS
\cite{Elsner:2008sn} results; small energy shifts among the different data sets are still 
possible. The data points in the forward region for incoming photon energies below 1~GeV 
(top row) are statistically limited and have very small degrees of polarization; thus, 
they have increased error bars. The beam asymmetries for the data set with a coherent peak 
position at 1610~MeV are shown in Fig.~\ref{Figure:Results1600} with the same energy 
binning. For energies above 1400~MeV, the data are extracted from the higher-energy data 
set alone. In the overlap region between 1200 and 1400~MeV, we have averaged the results 
from the two data samples (shown in Fig.~\ref{Figure:Results1600}) based on their good 
agreement.

The trigger conditions during the ``data taking'' were not optimized for the production 
of $\pi^0$~mesons over the full angular range. For this reason, the $\Sigma$~distributions 
exhibit a region of very low acceptance between about $65^\circ$ and $115^\circ$. Our 
acceptance cut of 8\,\% removes these data points. 

The results from this analysis are in excellent agreement with previous measurements.
Overall, the new photon beam asymmetries in the forward region and above 1500~MeV also agree 
nicely with the predictions of the SAID SP09 model~\cite{said:database,Dugger:2009pn}. However, 
small deviations are observed for energies above 1400~MeV, where the broad structure in 
the forward direction seems to underestimate the data for $\theta_{\rm c.m.} < 50^\circ$. 
The recently published solution of the Bonn-Gatchina PWA~\cite{Anisovich:2009zy} 
is in excellent agreement with the data and SAID over the full range of previously available 
data, but tends to systematically underestimate the data in the forward region. A new
solution including the results of this analysis is in preparation; the preliminary curve 
is given by the dashed-dotted line in Fig.~\ref{Figure:Results1350}. Small changes to the 
width and helicity couplings of the nucleon resonance $N(1720)P_{13}$ are observed. This is 
presently being investigated further and will be the subject of a forthcoming publication 
on {\it P-wave excited baryons} by the Bonn-Gatchina PWA group
\cite{Andrey:private}. A better understanding of the properties of the $N(1720)P_{13}$
resonance (from a coupled-channel analysis) will also help resolve its contribution to 
$\eta$~photoproduction. Its dominance over contributions from the $N(1710)P_{11}$ resonance 
to the reaction $\gamma p\to p\eta$ remains disputed.
 
The MAID\,2007 predictions (dashed curves)~\cite{Drechsel:2007if} show overall good agreement with 
SAID and the experimental data for energies below 1500~MeV. Significant deviations occur 
in the 932- and 965-MeV photon energy bins for central scattering angles
(Fig.~\ref{Figure:Results1350}). At photon energies greater than 1500~MeV, MAID\,2007 tends
to systematically underestimate the forward region and to overestimate the backward 
region because (precise) data have been missing. Our new results presented here and the 
recent LEPS data, which cover the backward region~\cite{Sumihama:2007qa}, will, thus, be useful to 
constrain future model solutions and partial wave analyses. 

Although it will be possible to modify the model solutions to better describe the data, 
double-polarization observables are needed to unambiguously extract the scattering 
amplitude.

\section{\label{Section:Summary}Summary}
To summarize, we have presented the results of a reanalysis of previously published 
CBELSA/TAPS data and new measurements of the beam asymmetry $\Sigma$ for the 
photoproduced $p\pi^0$ final state. New data points have been added to the very
forward direction of the $\pi^0$~meson in the c.m. system. The continuous
beam from the ELSA accelerator and the goniometer setup of the experiment provided
a linearly polarized tagged-photon beam for the coherent peak positions at 1305 
and 1610~MeV. The results are in very good agreement with the earlier measurements at
ELSA and also with previous results from other facilities.

\subsection*{Acknowledgments}
We thank all the participating institutions for their invaluable contributions to the 
success of the experiment. We acknowledge financial support from the National 
Science Foundation (NSF), Deutsche Forschungsgemeinschaft (DFG) 
within the SFB/TR16, and from Schweizerischer Nationalfond. The collaboration 
with St. Petersburg received funds from DFG and the Russian Foundation 
for Basic Research.

\end{document}